\documentclass{aa}
\input psfig.sty

\DeclareMathAlphabet{\mathsc}{OT1}{cmr}{m}{sc}
\def\testbx{bx}%
\DeclareRobustCommand{\ion}[2]{%
\relax\ifmmode
\ifx\testbx\f@series
{\mathbf{#1\,\mathsc{#2}}}\else
{\mathrm{#1\,\mathsc{#2}}}\fi
\else\textup{#1\,{\mdseries\textsc{#2}}}%
\fi}
\def\la{\mathrel{\mathchoice
{\vcenter{\offinterlineskip\halign{\hfil$\displaystyle##$\hfil\cr<\cr\sim\cr}}}
{\vcenter{\offinterlineskip\halign{\hfil$\textstyle##$\hfil\cr
<\cr\sim\cr}}}
{\vcenter{\offinterlineskip\halign{\hfil$\scriptstyle##$\hfil\cr
<\cr\sim\cr}}}
{\vcenter{\offinterlineskip\halign{\hfil$\scriptscriptstyle##$\hfil\cr
<\cr\sim\cr}}}}}

\def\ltsima{$\; \buildrel < \over \sim \;$}
\def\lsim{\lower.5ex\hbox{\ltsima}}
\def\gtsima{$\; \buildrel > \over \sim \;$}
\def\gsim{\lower.5ex\hbox{\gtsima}}

\begin{document}

\title{The discovery and study of  the optical counterparts of
the transient X--ray pulsars \object{RX\,J0052.1--7319} and
\object{XTE\,J0111.2--7317 in the SMC}
\thanks{Based on
observations carried out at ESO, La Silla, Chile (62.H--0513, 63.H--0294(B) and
64.H--0059(A))}}

\author{S.~Covino\inst{1} \and
I.~Negueruela\inst{2}  \and
S.~Campana\inst{1} \and
G.L.~Israel\inst{3, }\thanks{Affiliated to I.C.R.A.} \and
V.F.~Polcaro\inst{4} \and
L.~Stella\inst{3, \star\star} \and
F.~Verrecchia\inst{5}}

\offprints{S. Covino,
\email{covino@merate.mi.astro.it}}

\institute{Osservatorio Astronomico di Brera, Via Bianchi 46, I--23807 Merate (LC), Italy
\and Observatoire de Strasbourg, 11 rue de l'Universit\'e, F67000 Strasbourg, France.
\and Osservatorio Astronomico di Roma, Via Frascati 33, I--00040 Monteporzio Catone,
Roma, Italy
\and Istituto di Astrofisica Spaziale, Via Fosso del Cavaliere, I--00133 Roma, Italy
\and Universit\`a ``La Sapienza'', Dipartimento di Fisica, Piazzale A. Moro 5, I--00185, Roma, Italy}

\date{Received date / accepted date}

\titlerunning{RX\,J0052.1--7319 and XTE\,J0111.2--7317 optical counterparts}
\authorrunning{Covino et al.\ }

\abstract{
We report on the discovery and confirmation of the optical counterparts of the two
transient X--ray pulsars, \object{RX\,J0052.1--7319} and
\object{XTE\,J0111.2--7317}. In the narrow ($\sim$$3''$ radius) X--ray
error circle of \object{RX\,J0052.1--7319} we found a $V=14.6$ ($B-V=0.1$)
09.5IIIe (a classification as a B0Ve star is also possible, since the
luminosity class depends on the uncertainty on the adopted reddening).
Medium resolution spectra for this object show Balmer lines in emission with
an equivalent width of H$\alpha$=-12$\div$-16\,\AA.
In the X--ray error box of \object{XTE\,J0111.2--7317} we found a
relatively bright object ($V=15.4$, $B-V=0.06$) which has been classified as
a B0.5--B1Ve star and that was later confirmed by  Coe et al. (2000) as the most
plausible counterpart for \object{XTE\,J0111.2--7317}. Also in this case we easily detect Balmer emission lines
with an H$\alpha$ equivalent width of about $-21$\,\AA. There is also evidence
for the presence of a surrounding nebula, possibly a supernova remnant.
A further bright B0Ve star was found just outside the X--ray error circle of
\object{XTE\,J0111.2--7317}. We discuss the implication of these results in
the light of the current knowledge of Be/X--ray binary systems in the
Magellanic Clouds and within our Galaxy.
}

\maketitle


\section{Introduction}

X--ray pulsars in High Mass X--ray Binaries (HMXBs)  consist of an accreting neutron star
orbiting a (super)giant or a main--sequence Be--type companion star.
The greatest part of known neutron stars in Supergiant XBs (from now on SXBs)
show permanently high X--ray fluxes driven by accretion
of a roughly spherical dense wind from the massive companion (which may be
enhanced by Roche lobe overflow). In contrast, the neutron stars in BeXBs
often exhibit transient X--ray outbursts which may occur periodically at periastron
(type I) or when the companion star undergoes a mass loss episode from the
equatorial regions due to its high rotational velocity (up to $\sim$75\% of
the break--up velocity; type II). In BeXBs, the primary star is an
early type star in the range 10$\div$20 M$_\odot$ of luminosity class III to V, which
often displays Balmer lines, but also He\,I and metallic lines, in
emission. Due to these lines, Be stars are difficult
to classify. Moreover, since BeXBs are generally first detected as X--ray sources
with positional accuracies which can be as poor as a few arcminutes, the optical/IR
follow--up observations and the detection of the possible counterpart is often
a difficult task. In addition, the BeXB concentration on the Galactic Plane and
in the Magellanic Clouds (MCs) results in a further decrease in the chances
of detecting them because of crowding and source confusion.
So far only about $\sim$20 optical counterparts of BeXBs have been discovered
out of the $>$100 known and 10$^4$--10$^5$ expected Be/X--ray pulsars
(Nelson et al. \cite{NSW93} and \cite{NWSW95}).
Detailed discussions on the nature and properties of these systems have been
recently reported by several authors (Negueruela \cite{N98}; Coe \cite{C2000},
Negueruela \& Okazaki \cite{NH2000}).

In this paper we report on the results of optical observations aimed at
searching for the counterpart of two transient X--ray pulsars:
\object{RX\,J0052.1--7319} and \object{XTE\,J0111.2--7317}. Both systems are
hosted by the Small Magellanic Cloud (SMC) which harbors a large number
of these systems (Yokogawa et al. \cite{YO01}; Mereghetti \cite{M01}).

The paper is organized as follows: in Sect.\,\ref{sec:obs} we describe the
techniques we have applied for the observations and the data reduction and
in Sect.\,\ref{sec:todo} we discuss the
result of the observations and derive physical parameters for the proposed
optical counterparts of \object{RX\,J0052.1--7319} and
\object{XTE\,J0111.2--7317}. The detailed spectral classification of the
observed stars are also provided.

\section{Observations}
\label{sec:obs}

\begin{table*}
\caption{Record of the observations.}
\begin{center}
\begin{scriptsize}
\begin{tabular}{cccccccccc}
\hline
Date        & Telescope    & Field                       & Filter & Grism/grating  & Exposure  & Slit   & Range    & Resolution & Seeing   \\
            &              &                             &        & ESO \#         & (sec)     &(arcsec)& (\AA)    & (\AA)      & (arcsec) \\
\hline
1999 Jan 18 & 1.5m Danish  & \object{RX\,J0052.1--7319}  &   $V$  &                &\ 200      &        &          &            &     2.5  \\
1999 Jan 18 & 1.5m Danish  & \object{RX\,J0052.1--7319}  &   $R$  &                &\ 200      &        &          &            &     1.5  \\
1999 Jan 18 & 1.5m Danish  & \object{XTE\,J0111.2--7317} &   $V$  &                &\ 200      &        &          &            &     1.5  \\
1999 Jan 18 & 1.5m Danish  & \object{XTE\,J0111.2--7317} &   $R$  &                &\ 150      &        &          &            &     2.0  \\
1999 Jan 19 & 1.5m Danish  & \object{RX\,J0052.1--7319}  &   $B$  &                &\ 200      &        &          &            &     3.0  \\
1999 Jan 18 & 1.5m Danish  & \object{XTE\,J0111.2--7317} &   $B$  &                &\ 200      &        &          &            &     3.5  \\
1999 Jan 19 & 1.5m Danish  & \object{RX\,J0052.1--7319}  &        & \ 4            & 1000      &  2.0   &3500--9000& 15         &     1.7  \\
1999 Jan 19 & 1.5m Danish  & \object{XTE\,J0111.2--7317} &        & \ 4            & 1000      &  2.0   &3500--9000& 15         &     1.8  \\
1999 Jan 20 & 1.5m Danish  & \object{RX\,J0052.1--7319}  &        & \ 8            & 2000      &  1.5   &5800--8300& \ 5        &     1.7  \\
1999 Jan 20 & 1.5m Danish  & \object{XTE\,J0111.2--7317} &        & \ 8            & 2000      &  2.0   &5800--8300& \ 6        &     1.6  \\
1999 Sep 14 & 3.6m ESO     & \object{XTE\,J0111.2--7317} &        & \ 7            &\ 900      &  1.0   &3300--5200& \ 6        &     1.0  \\
1999 Sep 14 & 3.6m ESO     & \object{XTE\,J0111.2--7317} &        & \ 9            &\ 900      &  1.0   &4700--6800& \ 6        &     1.0  \\
1999 Sep 15 & 3.6m ESO     & \object{RX\,J0052.1--7319}  &        & \ 7            & 2400      &  1.0   &3300--5200& \ 6        &     1.1  \\
1999 Sep 15 & 3.6m ESO     & \object{RX\,J0052.1--7319}  &        & \ 9            & 2400      &  1.0   &4700--6800& \ 6        &     1.1  \\
1999 Nov \ 2& 1.5m ESO     & \object{XTE\,J0111.2--7317} &        & 33             & 1500      &  2.0   &3500--6000& \ 3        &     1.5  \\
\hline
\end{tabular}
\end{scriptsize}
\end{center}\label{tab:log}
\end{table*}

The search for the optical counterparts in the fields of the X--ray sources
\object{RX\,J0052--7319} and \object{XTE\,J0111.2--7317} was carried out by means
of several techniques and instruments. Photometry was performed in order to
study the stellar field main properties and to select objects with colors
and magnitudes expected for the counterparts of a BeXB in the SMC.
Spectroscopy of the optical counterpart candidates was carried out to infer
the physical parameters (i.e. spectral classification) and to look for signatures
of activity to be correlated with the X--ray emission.

Multicolor photometry ($B$, $V$, and $R$ bands) was carried out for each field with the
1.5m Danish telescope of the ESO at La Silla (Chile). Instrumental magnitudes
for each stellar object in the images were derived with aperture and
profile--fitting photometry by means of the DAOPHOT\,II package (Stetson
\cite{S87}). Low and medium resolution spectroscopy was also performed at
the 1.5m Danish and 3.6m telescopes of the ESO at La Silla for a
number of selected bright objects in the X--ray error circle. When the observing conditions were adequate
we also performed absolute calibration both for photometry and spectroscopy.
Spectroscopic data were analyzed by the ``long'' context in the ESO--MIDAS
package (96NOV and later versions). Medium resolution spectroscopy of
\object{XTE\,J0111.2--7317} was also obtained with the ESO 1.5m telescope. These
observations were reduced with the {\sc Figaro} package (Shortridge et al.
\cite{SMC97}) of the {\em Starlink} suite. The S/N of the spectra obtained
at the 3.6m was always of the order of 150 (or better) while those obtained at
the ESO 1.5m had a S/N around 50. The spectral features discussed in the text
were present in at least two spectra or clearly above the noise level.

A full record of the optical observations is reported in Table \ref{tab:log}.

\section{Discussion}
\label{sec:todo}

\subsection{\object{RX\,J0052.1--7319}}
\label{sec:aretha}

The X--ray transient \object{RX\,J0052.1--7319} was discovered by Lamb et al.
(\cite{L81}) with the analysis of ROSAT HRI and BATSE data. The object showed a
period of 15.3\,s (Kahabka \cite{K98} and \cite{K99}) and a flux in the
0.1--2\,keV band of $2.6 \times 10^{-11}$\,erg\,s$^{-1}$\,cm$^{-2}$.

We devoted particular care  to the definition of the X--ray error circles of this source. The
archival ROSAT HRI images, available for \object{RX\,J0052--7319}, were
analyzed using both a sliding cell and a wavelet transform--based algorithm
(Lazzati et al. \cite{LCR99} and Campana et al. \cite{CLP99}). These
techniques, already applied by us to the ROSAT HRI fields of several X--ray
sources (Israel et al. \cite{ICS99}, Israel et al. \cite{ICC00}, Covino et al.
\cite{CIP00}, Israel et al. \cite{ICP00}, Israel et al. \cite{IPC00},
Mereghetti et al. \cite{MMC00}), allowed us to infer a positional accuracy in
the $5 \div 10''$ range.
In particular \object{RX\,J0052.1--7319} was observed three times in the
BMW--HRI source catalog (see Table \ref{tab:xlog}). The source is highly
variable with a $\gsim$1,000 variation in the count rate. The source is
serendipituously detected at relatively large off--axis angles ($>8'$) affecting
the source positioning. For each field we are able to perform a boresight
correction, registering the X--ray source positions to an optical frame. In the
case of the shortest observation, we did not find
X--ray sources suitable for the boresight. For the remaining two observations
we used for the boresight correction 3 and 1 sources, respectively.
Boresight errors are summed up in quadrature with the
source location error (see Table \ref{tab:xlog}). We then combined the source
positions obtained in the three observations to derive the final position:
R.A.= 00$^{\rm h}$ 52$^{\rm m}$ 13$^{\rm s}$.27, DEC.= --73$^{\rm o}$ $19'$ $19.5''$
(J2000). The 90\% error circle radius amounts to $3.3'$.

\begin{table*}
\caption{Source characteristics of \object{RX\,J0052.1--7319} in the BMW--HRI
catalog.}
\begin{center}
\begin{scriptsize}
\begin{tabular}{ccccccc}
\hline
Observation & RA (J2000)  & DEC (J2000) & Error       & Off--axis &Exp. time & Count rate  \\
  (ROR)     &             &             & ($''$, 90\%)& (armin)  & (s)      & (c s$^{-1}$)\\
\hline
rh300513n00 & 00 52 12.60 & --73 19 19.7&  10.00      & 20.9     & \ 1582.4 & 0.9979      \\
rh600811n00 & 00 52 14.41 & --73 19 19.3&\  2.07      &\ 8.4     &  27395.0 & 0.0009      \\
rh600812a01 & 00 52 13.10 & --73 19 19.9&\  2.53      & 16.7     &  17064.5 & 0.6863      \\
\hline
\end{tabular}
\end{scriptsize}
\end{center}\label{tab:xlog}
\end{table*}

\begin{figure}[htb]
\centerline{\psfig{figure=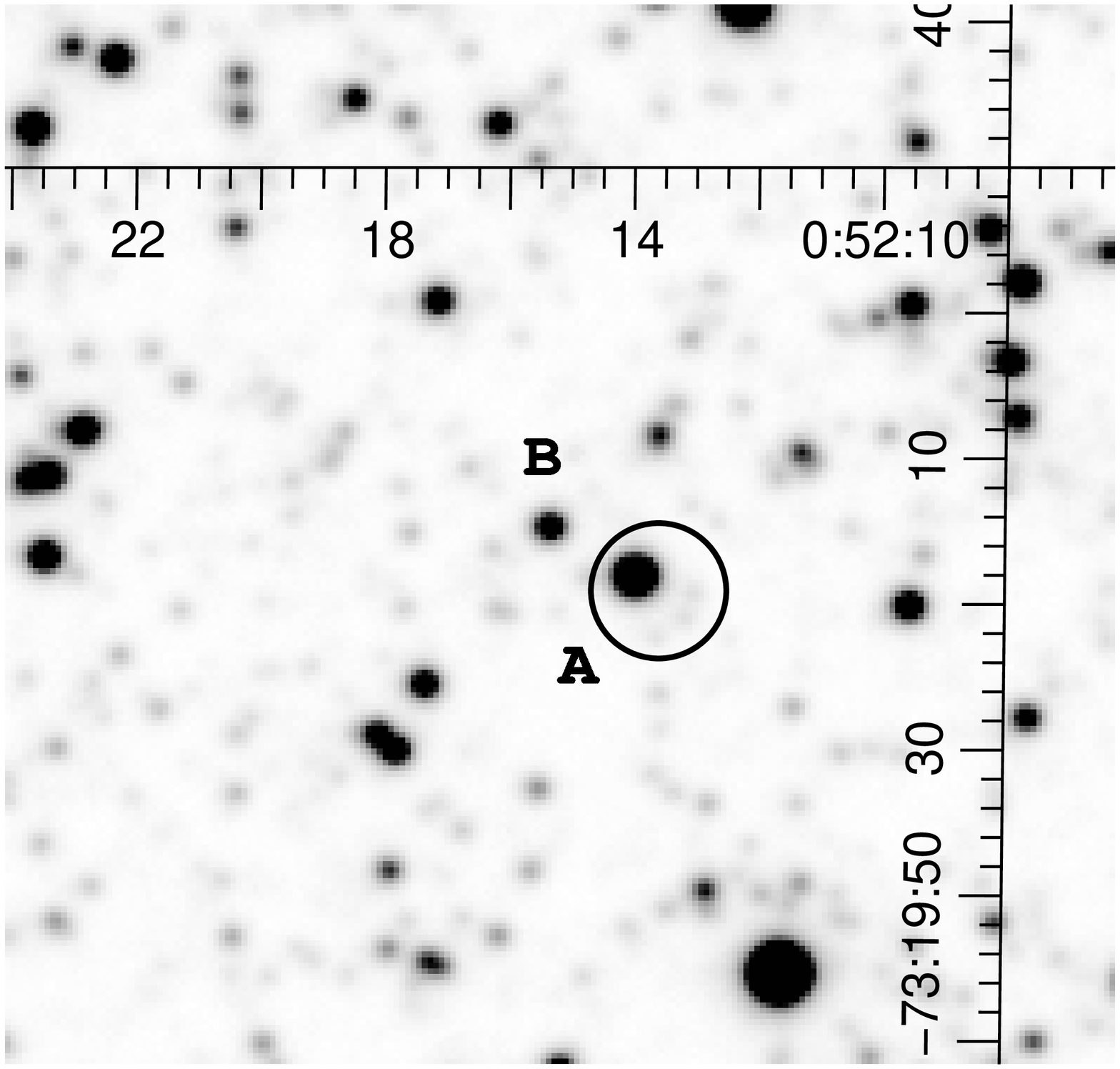,width=8cm} }
\caption{1.5m ESO Danish telescope \object{RX\,J0052--7319} $R$ image. The X--ray
error circle is also shown with the optical counterpart labeled as A. A nearby
object is labeled as B.}
\label{fig:aretha_opt}
\end{figure}

An optical counterpart for this source was proposed (Israel et al.
\cite{ISCCM99}), a B--type star with $R = 14.54 \pm 0.03$ and $V-R = +0.08
\pm 0.04$, located at coordinates (J2000, estimated uncertainty
$1''$) R.A. = $00^{\rm h}$ $52^{\rm m}$ $13.9^{\rm s}.0$, DEC.=
$-73^{\rm o}$ $19'$ $19''$. In Fig.\,\ref{fig:aretha_opt} a $R$ frame for the
\object{RX\,J0052.1--7319} field is shown, together with the X--ray
error--circle.
The proposed counterpart is the object labeled as A. This
star was also identified in the OGLE photometric database (labeled as
\object{SMC\_SC6\,99923}, Udalsky et al. \cite{U99}) as a long--term variable star
with quasiperiodic light variation of amplitude 0.13\,mag in the I band.
This optical object is within the X--ray error circle of
\object{RX\,J0052.1--7319}, lying at $2.8''$ from the best X--ray
position.


\begin{figure}[bth]
\centerline{\psfig{figure=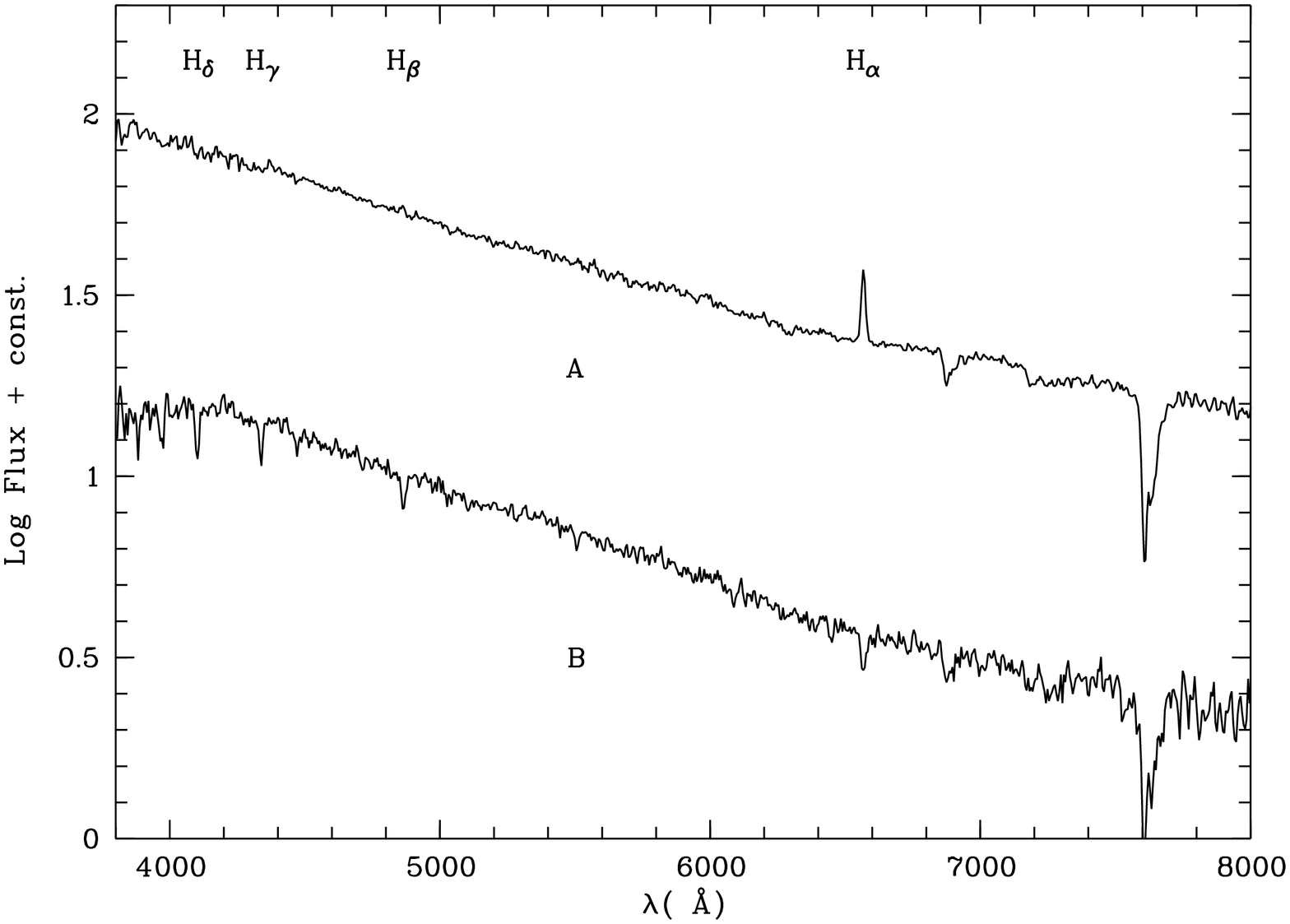,width=9cm,height=7cm,angle=-90} }
\caption{1.5m ESO Danish telescope \object{RX\,J0052--7319} spectrum for objects
A and B. Position of the main hydrogen lines is also shown.}
\label{fig:aretha_specD1}
\end{figure}

A 1,000 s optical spectrum at the 1.5m Danish telescope of the ESO at La Silla
(Fig.\,\ref{fig:aretha_specD1}) for object A shows H$\alpha$ and
H$\beta$ emission (EW $\sim -12$ and $-3$\,\AA, respectively), supporting the Be
nature of the star.  Star B, just outside the X--ray error circle, does not show any
emission line. Its spectrum is compatible with being a moderately bright young
SMC star somewhat later than object A.

\begin{figure*}[bth]
\centerline{\psfig{figure=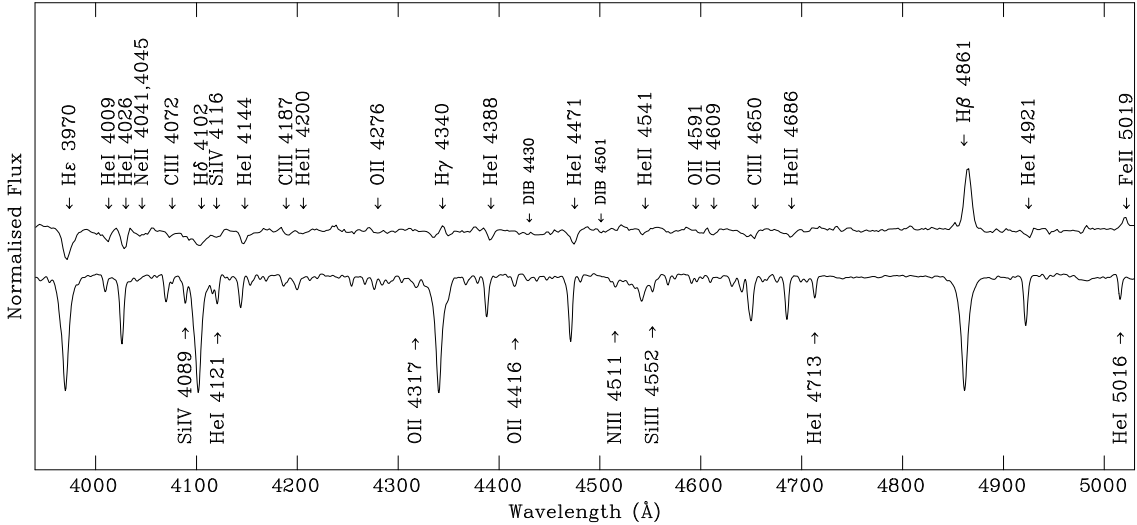,width=10cm} }
\caption{Spectrum of the optical counterpart of \object{RX\,J0052--7319} in the
classification region compared to the B0V standard $\nu$ Ori. Note the broadness
and shallowness of all features in \object{RX\,J0052--7319}. Some relatively
prominent lines not seen in the spectrum of \object{RX\,J0052--7319} are marked
on the spectrum of $\nu$ Ori. Both spectra have been divided by a spline fit to
the continuum for normalization and smoothed with a $\sigma=1.2$ Gaussian
filter. The identification of \ion{O}{ii} lines is tentative, since they do not
correspond well to the derived spectral type, and the features could be notches
created by the presence of weak \ion{Fe}{ii} emission lines.}
\label{fig:aretha}
\end{figure*}

The spectrum of the optical counterpart of \object{RX\,J0052--7319} in the
classification region was also studied with the 3.6m ESO telescope and it is
displayed in Fig.\,\ref{fig:aretha}, together with that of the B0V MK
standard $\nu$ Ori. H$\beta$ and H$\gamma$ are in emission and display
asymmetric profiles. Some of the weaker \ion{He}{i} lines are not visible,
presumably filled in by emission. All the lines in the spectrum are very
shallow and broad, which is typical of a Be/X--ray binary counterpart. The
presence of weak \ion{He}{ii} lines places the object close to B0.

If the object is on the main sequence, the presence of weak
\ion{He}{ii}~$\lambda4200$\,\AA\ and the condition
\ion{He}{ii}~$\lambda4541$\,\AA~$\>>$~\ion{Si}{iii}~$\lambda4552$\,\AA\
give a spectral type B0Ve. For this spectral type, an intrinsic
$(B-V)_{0}=-0.26$ is expected (Wegner \cite{W94}). Therefore,
the measured $(B-V)=0.12$ implies $E(B-V)=0.38$. Such reddening is
rather high for a source in the SMC and would suggest that there is an
important local contribution to the reddening. It is difficult to
quantify if this could be due to localized absorption or to circumstellar
reddening due to emission from the disc. The relationships of Fabregat \&
Reglero (\cite{FR90}) would indicate for an H$\alpha$ EW= $-16$\,\AA\ the
circumstellar contribution should only be $E^{\rm cs}(B-V)\simeq0.04$, but
this relationship is only valid 
for isolated Be stars and the envelopes of Be/X-ray binaries are known to be 
denser than those of isolated stars (Zamanov et al. \cite{Z01}) -- therefore 
the circumstellar contribution to the reddening could be higher. 

The lower limit $E^{\rm cs}(B-V)\simeq0.04$ would result in 
$M_{V}=-5.18\pm0.15$, using  the distance
modulus $(M-m)=18.75\pm0.07$ (Udalski \cite{U00}). This intrinsic 
magnitude is far too bright for a main--sequence object and would
suggest that the star is a giant. Alternatively, we can assume that
the interstellar reddening takes a value $E_{\rm is}(B-V)=0.08$
typical for the SMC and that the rest of the reddening is
circumstellar, in which case $M_{V}=-4.4\pm0.15$. This value is  
compatible with a B0V star with $E_{\rm cs}(B-V)=0.3$, though we note
that such circumstellar emission would be rather high. 
Even though the weakness of all the Si lines
(\ion{Si}{iv}~$\lambda4089$\,\AA\ is hidden on the blue wing of
H$\delta$) would support a main sequence classification for a Galactic
object, we note that the metal content of the SMC is much lower than
that of the Milky Way and that even
supergiants in the SMC can have metallic lines as weak as those of 
main--sequence objects in our Galaxy (Walborn \cite{W83}; see also Lennon 
\cite{L97}). As a consequence, we find that we cannot decide on the
luminosity class of the star without knowing how much of the
reddening is circumstellar. If it is a giant, the presence of weak
\ion{He}{ii}~$\lambda4200$\,\AA\ would indicate spectral type
O9.5IIIe, while B0Ve would be the spectral classification if it is on
the main sequence.

\subsection{\object{XTE\,J0111.2--7317}}
\label{sec:eschilo}

The X--ray transient \object{XTE\,J0111.2--7317} was discovered by the RXTE
X--ray observatory in November 1998 (Chakrabarty et al. \cite{CLC98}). Analysis
of ASCA observation (Chakrabarty et al. \cite{COPY98}, Yokogawa et al.
\cite{YBM00})  identified this source as as a 31 s X--ray pulsar with a flux in
the $0.7 \div 10$\,keV band of $3.6 \times 10^{-10}$\,erg\,s$^{-1}$\,cm$^{-2}$
and $\sim$45\% pulsed fraction. The detection was also confirmed from the BATSE
telescope on the CGRO satellite which detected the source in the hard $20 \div
50$\,keV band with a flux ranging from 18 to 30\,mCrab (Wilson \& Finger
\cite{WF98}).

The best X-ray position for \object{XTE\,J0111.2--7317} was inferred from
the ASCA observation carried out during the 1998 outburst (Yokogawa et al.
\cite{YBM00}). Moreover, we applied the new calibration for the restoration of
the ASCA source position accuracy applying the algorithm described by Gotthelf
et al. (\cite{GUFKY00}). This yielded the following improved position of the
source for the ASCA--SIS observation: R.A.=1$^{\rm h}$ 11$^{\rm m}$ 11.5$^{\rm
s}$.5, DEC.=--73$^{\rm o}$ $16'$ $44.6''$ (J2000),  with an error radius of
$15"$ (90\% confidence). As a reference,  we also report the position as
inferred in the analysis of the ASCA--GIS observation: R.A.=1$^{\rm h}$ 11$^{\rm
m}$ 12.3$^{\rm s}$.5, DEC.=--73$^{\rm o}$ $16'$ $47.6''$ (J2000),  with an error
radius of $25"$ (90\% confidence).

No ROSAT observations were able to detect this source. Summing up all the 6
available PSPC observations we are able to derive a 3$\,\sigma$ upper limit of
$1.4\times 10^{-3}$ c s$^{-1}$ (0.1--2.4 keV, for a total exposure time of
$\sim$90 ks), which translates to a luminosity limit of $\sim 3\times 10^{34}$
erg s$^{-1}$ in the $0.1 \div 2.4$\,keV energy band for a Crab--like spectrum
and a column density of $10^{21}$ cm$^{2}$. These observations clearly indicate
the transient nature of this source.

\begin{figure}[hbt]
\centerline{\psfig{figure=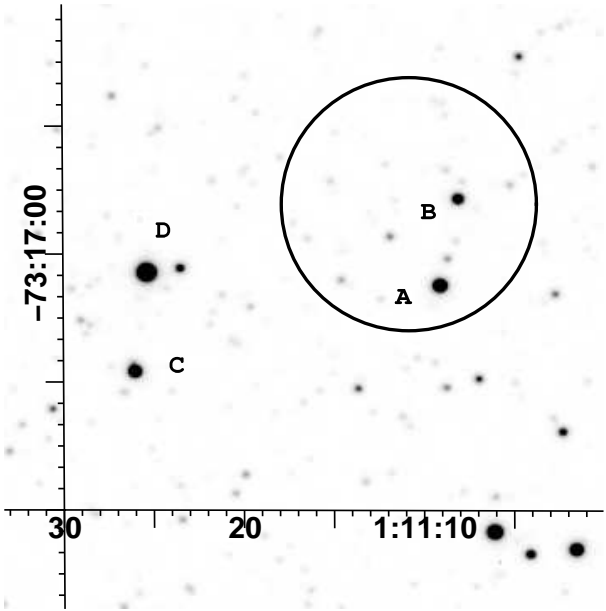,width=8cm} }
\caption{1.5m ESO Danish telescope \object{XTE\,J0111.2--7317} $R$ image.
The refined ASCA--SIS  X--ray error circle is also shown. The two brightest stars in the error
circle are labeled A and B; while the other two bright stars just outside the
error circle are labeled as C and D.}
\label{fig:eschilo_opt}
\end{figure}

A Be star counterpart was proposed by Israel et al. (\cite{ISCCM99}) and
subsequently confirmed by Coe et al. (\cite{CHR00}) as a B0--B2 star with strong
evidence for the presence of a surrounding nebula, possibly a SuperNova Remnant
(SNR).

Images for this field were taken (see Fig.\,\ref{fig:eschilo_opt})
revealing two bright stars within the $15''$ error circle. The fainter one
(star B) is a B--type star ($R = 15.29 \pm 0.03, V-R = +0.06 \pm 0.04$), located
at the coordinates (J2000, estimated uncertainty $1''$) R.A.= 01$^{\rm h}$
11$^{\rm m}$ 08$^{\rm s}.4$, DEC.= $-73^{\rm o}$ $16'$ $46''$.


A 1,000\,s spectrum taken with the 1.5m Danish telescope (Table \ref{tab:log})
revealed strong H$\alpha$ and H$\beta$ emission lines (EW $\sim -21$ and
$-1.5$\,\AA,
respectively), indicating that this source is likely the counterpart of the
X--ray transient. The brighter star ($R = 14.30 \pm 0.03$) does not show any
emission line. No other object brighter than $R \sim 17.3$ are found within the
error circle. One more B--type star ($R = 14.55 \pm 0.03, V-R = +0.05 \pm
0.04$), located at the coordinates (J2000, estimated uncertainty
$1''$) R.A.= $01^{\rm h}$ $11^{\rm m}$ $25^{\rm s}.9$, DEC.=
$-73^{\rm o}$ $17'$ $27''$ ($~1'$ away from the X--ray position, star C)
 also shows strong H$\alpha$ and H$\beta$ emission lines (EW = $-36$ and
$-3$\,\AA, respectively).

In order to infer an accurate spectral classification of the selected objects we
observed again the field and, on the 1999 September 15th, we set
the slit so that the bright Be star just outside the error circle (star C) was
also included in the slit. This star can be seen on the Eastern edge of all the
images in Coe et al. (\cite{CHR00}) and in our Fig.\,\ref{fig:eschilo_opt}. The
orientation of the slit was therefore almost E--W. On 1999 November 1st, on the
other hand, we set the slit so that the core of the dark nebulous region that
can be seen In Fig.\,2 of Coe et al. (\cite{CHR00}) to the SW of star B was
inside the slit.

\begin{figure*}[hbt]
\centerline{\psfig{figure=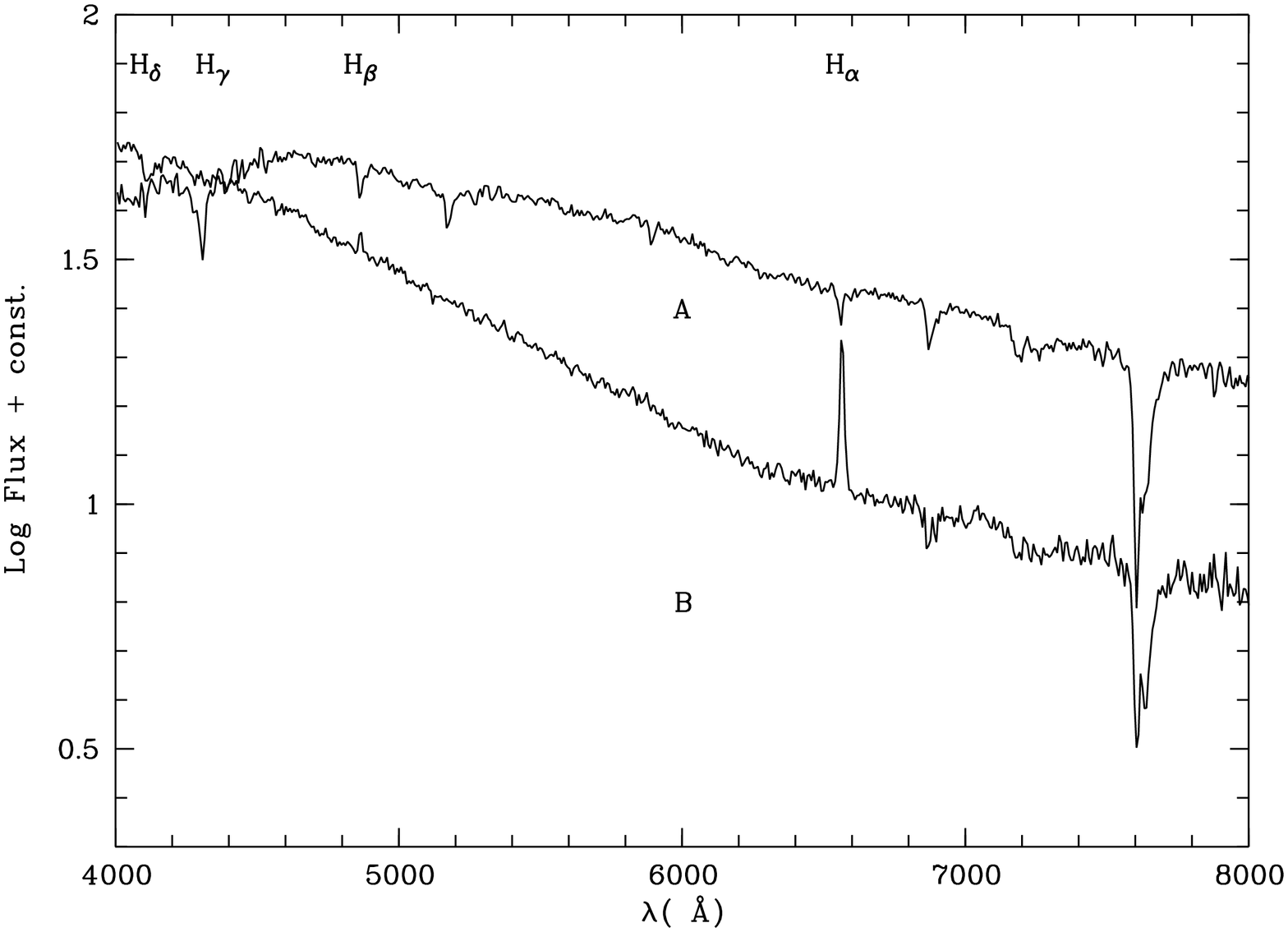,width=9cm,height=7cm,angle=-90}
 \psfig{figure=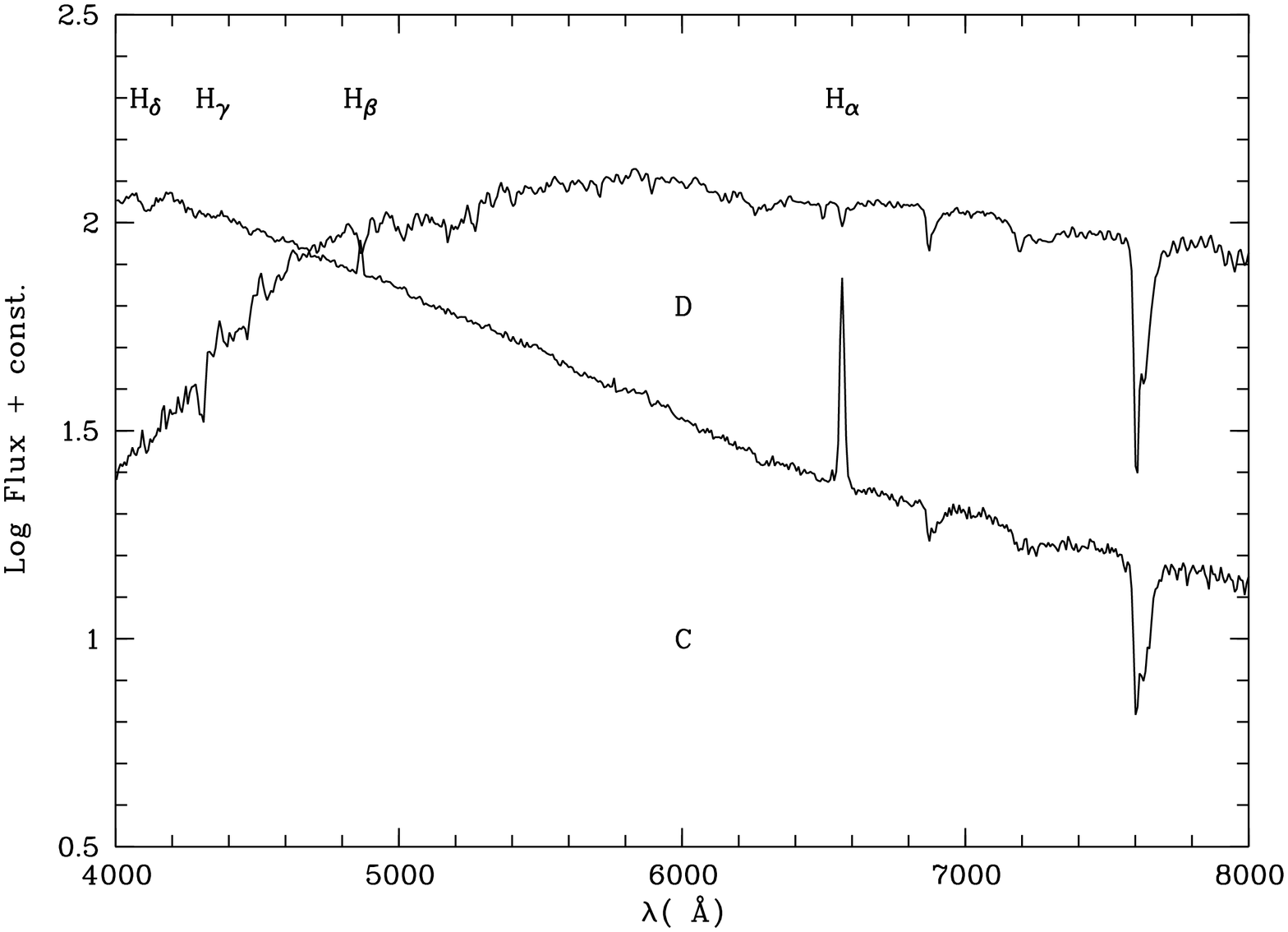,height=7cm,width=9cm,angle=-90}}
\caption{1.5m ESO Danish telescope \object{XTE\,J0111.2--7317} spectrum for
objects A, B (left panel), C and D (right panel) in the X--ray error
circle. Position of the main hydrogen lines is also shown.}
\label{fig:eschilo_specD1}
\end{figure*}

\begin{figure*}
\centerline{\psfig{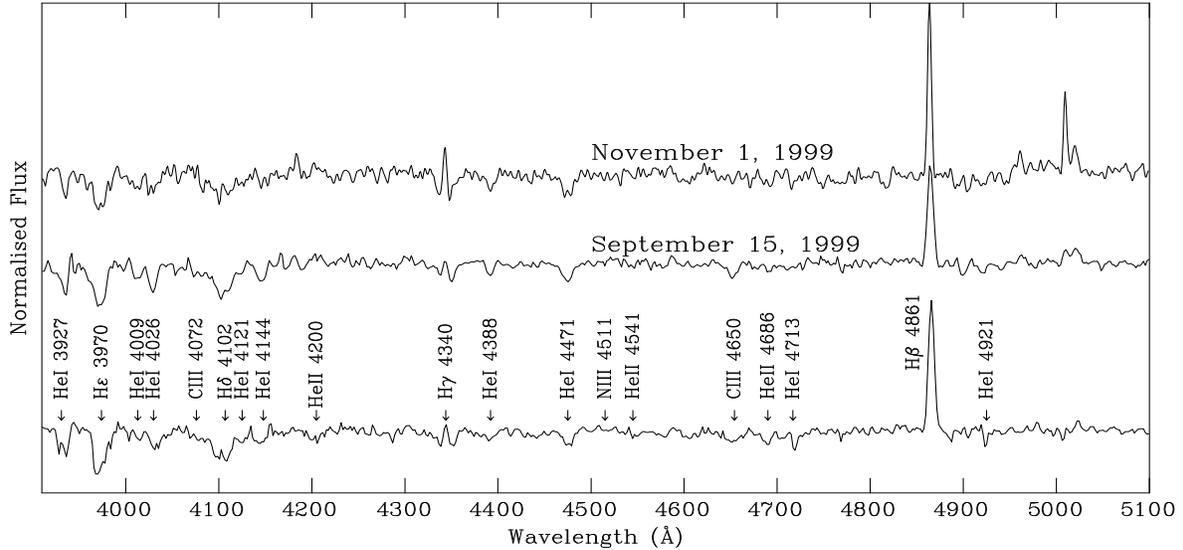} }
\caption{The two blue spectra of the optical counterpart to \object{XTE\,
J0111.2--7317}
(top) compared to that of the B0Ve star just outside the error circle (star C,
bottom). The lines marked on the spectrum of star C are characteristic of the
spectral type (compare with the spectroscopic standard in Fig.\,\ref{fig:aretha}).
All spectra have been divided by a spline fit to the continuum for normalization and
smoothed with a $\sigma=1.2$ Gaussian filter. All identified
features are present in at least two spectra.}
\label{fig:eschilo1_spec}
\end{figure*}

\begin{figure*}
\centerline{\psfig{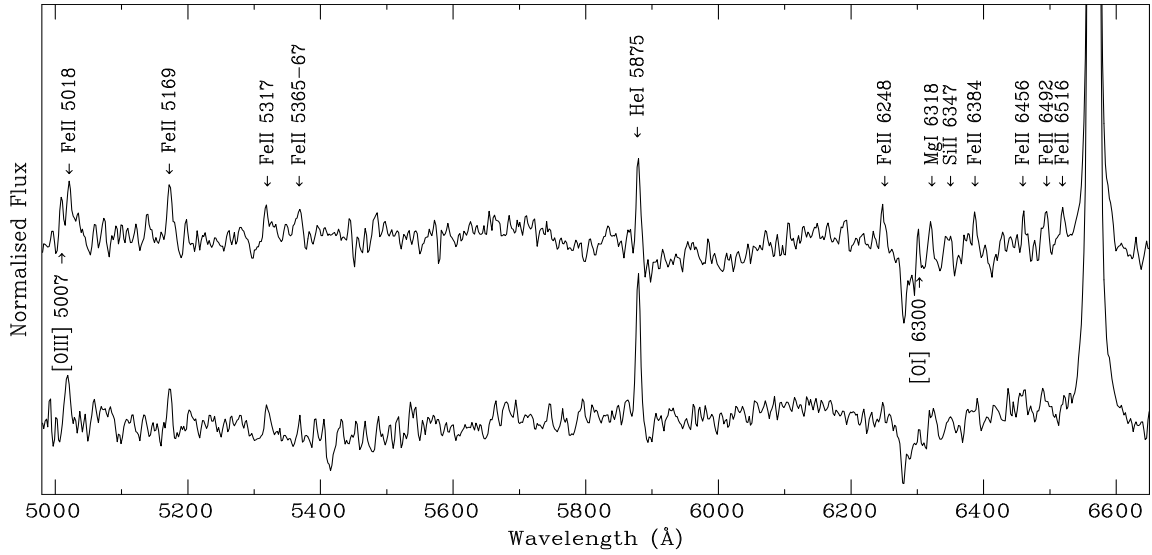} }
\caption{Red spectra of the optical counterpart to \object{XTE\,J0111.2--7317}
(top) and the B0Ve star just outside the error circle (star C, bottom). The
many permitted emission lines on the spectrum of the optical counterpart to
\object{XTE\,J0111.2--7317} indicate that it is intrinsically a Be star, while
the presence of weak forbidden lines must be due to emission from the
surrounding nebula. Both spectra have been divided by a spline fit to the
continuum for normalization and smoothed with a $\sigma=1.2$ Gaussian filter.}
\label{fig:eschilo2_spec}
\end{figure*}

The two spectra of star B in the classification region are displayed in
Fig.\,\ref{fig:eschilo1_spec}. The spectrum from November 1st has a higher
resolution, but the spectrum from September 15th has rather higher SNR. Also
displayed is the spectrum of star C, which shows H$\beta$ in emission and
emission components in other Balmer series lines. The measured H$\beta$
EW= $-3.3$\,\AA\ is typical of Be stars with strong emission. This is confirmed
by the red spectrum of the source (see Fig.\,\ref{fig:eschilo2_spec}), where
H$\alpha$ is strongly in emission (EW= $-40$\,\AA). Typically, Be stars with such
strong emission in the Balmer series also display many weak emission lines
corresponding to low--ionization metallic transitions (mainly \ion{Fe}{ii}). The
emission spectrum of star C is clearly seen in Fig.\,\ref{fig:eschilo2_spec}.

The blue spectrum of star C shows weak \ion{He}{ii} lines, indicating a
spectral type earlier than B0.5. The weakness of all metallic lines (and
specifically the \ion{Si}{iv} lines in the wings of H$\delta$) suggests a main
sequence object (though see previous section). The weak
\ion{He}{ii}~$\lambda\lambda$ 4200, 4541\,\AA\
indicate a spectral type O9.5--B0, while
\ion{He}{ii}~$\lambda4686$\,\AA$\la$~\ion{He}{i}~$\lambda4713$\,\AA\ favors the
latter classification. Assuming a distance modulus $(m-M)=18.8$ (Coe et al.
\cite{CHR00}), the measured $V=14.6$ implies a $M_{V}=-4.2$, consistent with a
B0V star (Vacca et al. \cite{VGS96}). Considering an interstellar reddening
$\approx 0.14$ (Coe et al. \cite{CHR00}), the source must actually have
$M_{V}=-4.6$, which is slightly too high for a main sequence object. However, Be
stars tend to be brighter than that corresponds to their spectral type due to
the continuum emission from the circumstellar disc. Zorec \& Briot (\cite{ZB91})
and Fabregat \& Torrej\'{o}n (\cite{FT98}) give an average difference of 0.3 mag
between Be stars and main--sequence objects. Using the approximate relationship
of Fabregat \& Reglero (\cite{FR90}), which is valid for isolated Be stars, an
H$\alpha$ EW= $-40$\,\AA\ implies a circumstellar brightening $\Delta V \simeq
0.3$. Therefore star C has a spectral type (B0Ve) and spectral morphology
typical of a counterpart to Be/X--ray binary.

The spectrum of star B, on the other hand, does not show obvious
\ion{He}{ii} lines, indicating a later spectral type. The only photospheric
features that could be visible in its spectrum and are not marked on the
spectrum of star C are the \ion{O}{ii} blends at
$\lambda\lambda$~$4146-53$ and $4185-90$\,\AA, \ion{He}{i}~$\lambda4169$\,\AA\ and
perhaps \ion{N}{ii}~$\lambda4237-4242$\,\AA. These are possible identifications
for features that are visible in both spectra of the source. However, from the
two spectra, it is clear that longwards of $\approx \lambda4200$\,\AA\
the observed ``continuum'' actually consists of a forest of weak emission lines,
that create apparent absorption features, such as the shallow band at
$\sim \lambda4600$\,\AA\ in the November spectrum, which is probably only a gap
between emission components centered on \ion{Fe}{ii}~$\lambda4584$\,\AA\ and
\ion{Fe}{ii}~$\lambda4629$\,\AA.

If the weak feature seen at $\approx \lambda4689$\,\AA\ in both spectra is
real and can be identified with \ion{He}{ii}~$\lambda4686$\,\AA\ (with the
radial velocity shift of $165\:{\rm km}\,{\rm s}^{-1}$ measured by Coe et al.
\cite{CHR00}), then the spectral type would be B0.5--B0.7. The presence of the
strong \ion{O}{ii}/\ion{C}{iii} blend at $\approx \lambda4650$\,\AA\ indicates
that it cannot be later than B1 in any case.

Longwards of H$\beta$ the \ion{Fe}{ii} emission spectrum is very strong (see
Fig.\,\ref{fig:eschilo2_spec}). Close to $\lambda5,000$\,\AA, apart from
\ion{Fe}{ii}~$\lambda5018$\,\AA, two other lines, which are particularly
strong in the November spectrum, can be seen. Their wavelengths suggest
identification with the $\left[\ion{O}{iii}\right]$~$\lambda\lambda$~4959 and
5007\,\AA\ lines. Such lines are not generally seen on stellar spectra, but are
typical of nebular emission.

Diffuse emission is actually apparent on the region immediately to the
East of star B in our September image. When the sky spectrum from that
area is extracted and a sky spectrum from a distant area is subtracted, strong
emission lines corresponding to $\left[\ion{O}{ii}\right]$~$\lambda3727$\,\AA,
H$\alpha$, H$\beta$, H$_\gamma$,
$\left[\ion{O}{iii}\right]$~$\lambda\lambda$~4959, 5007\,\AA\ and
presumably $\left[\ion{S}{ii}\right]$~$\lambda6716$\,\AA\ can be seen.

In the image corresponding to the November observation, diffuse emission
is present  on the region immediately to the SouthWest of star B.
$\left[\ion{O}{ii}\right]$~$\lambda3727$\,\AA, H$\beta$, and
$\left[\ion{O}{iii}\right]$~$\lambda\lambda$~4959, 5007\,\AA\ are readily
seen in the extracted sky spectrum.

Extracting several sky spectra around star B, we come to the conclusion
that most of the nebular line emission comes from the immediate vicinity of the
star and that it is strongest on the spatial position of the star.
Therefore we conclude that a significant portion of the emission seen
on star B comes from the nebulosity surrounding it and we are tempted
to attribute (at least part of) the obvious differences between the
September and November spectra to the different slit orientations.

However, it is clear from the detection of many permitted \ion{Fe}{ii}
emission lines that star B is also intrinsically an emission line star,
i.e., is surrounded by a circumstellar envelope. As a matter of fact, we believe
that star B is also a Be star, since other possibilities seem to be ruled
out because of the following points: ({\bf i}) the detected forbidden lines are
typical of high excitation nebulae and not of B[e] stars. ({\bf ii}) The presence
of a compact companion precludes the possibility that star B is a Herbig Be star.
({\bf iii}) The EWs of H$\alpha$ and H$\beta$ are typical of Be stars and very
low for any B[e] star. ({\bf iv}) If star B was a B[e] star, it should show
narrow emission lines corresponding to [\ion{Fe}{ii}]. Moreover, the $E(B-V)=0.2-0.3$
derived by Coe et al. (\cite{CHR00}) shows that the obscuration by the nebula
is small. Therefore the luminosity of star B is far too low for a B[e]
star.

Therefore star B has a spectral type B0.5--B1Ve, compatible with the
$M_{V}=-3.8$ derived by Coe et al. (\cite{CHR00}) and it is the likely
counterpart to \object{XTE\,J0111.2--7317}. Its association with the nebulosity
that apparently surrounds it is almost certain, though further study is
necessary in order to assess its nature.
\begin{table*}[tbh]
\caption{Photometry for selected objects in the fields of
\object{RX\,J0052.1--7319} and \object{XTE\,J0111.2--7317}. The proposed
counterparts are labeled by a ``*''. Errors are between $0.03 \div 0.05$\,mag.}
\begin{center}
\begin{tabular}{llcccl}
\hline
Field                       & Star  & $B$     & $V$     & $R$     & Spectral type\\
\hline
\object{RX\,J0052.1--7319}  & A*    & 14.73 & 14.62 & 14.54 & O9.5IIIe     \\
                            & B     & 15.82 & 15.92 & 16.05 &              \\
\object{XTE\,J0111.2--7317} & A     & 15.32 & 14.65 & 14.30 &              \\
                            & B*    & 15.42 & 15.36 & 15.29 & B0.5--B1Ve   \\
                            & C     & 14.63 & 14.61 & 14.55 & B0Ve         \\
                            & D     & 15.05 & 13.67 &       &              \\
\hline
\end{tabular}
\label{tab:photo}
\end{center}
\end{table*}

We note that our value $(B-V)=0.05$ would imply an $E(B-V)=0.28$, consistent
with the value derived by Coe et al. (2000) from Str\"{o}mgren photometry
(though hardly consistent with the value they derived from Johnson photometry,
which has $(B-V)=-0.08$. However, the two Coe at al. measurements of the $V$
magnitude reported for this star show a spread that is much larger than the
measurement errors, indicating variability.

\section{Conclusions}

We discuss the optical conterparts of two transient X--ray pulsars in the SMC.
In the case of \object{RX\,J0052.1--7319}, ROSAT HRI observations allowed us
to derive an error circle of $\sim$$3''$ radius (90\%). Accurate
positioning can then be achieved with past HRI instruments if a boresight
correction can be derived. Within the error circle only one relatively bright
source is found with $V=14.6$ ($B-V=0.1$). The object has been investigated
photometrically and spectroscopically leading to classify it as a 09.5IIIe star
(a classification as a B0Ve star is also possible, since the luminosity class
depends on the adopted reddening). Medium resolution spectra show prominent
H$\alpha$ and H$\beta$ emission lines, further strengthening the identification.

The position of \object{XTE\,J0111.2--7317} is based on ASCA observations
analyzed by the new calibration for the restoration of the ASCA source position
accuracy described by Gotthelf et al. (\cite{GUFKY00}).  The analysis provided
 a $15''$ error box. Within it we found a relatively bright object
($V=15.4$, $B-V=0.06$) which can be classified as a B0.5--B1Ve star. Also in
this case we easily detect Balmer emission lines. A further bright B0Ve star was
found outside the X--ray error circle. Coe et al. (\cite{CHR00}) found
evidence for the presence of a surrounding nebula, possibly a supernova remnant,
around the likely counterpart of \object{XTE\,J0111.2--7317}.


\begin{acknowledgements}
SCo thanks the Rome Astronomical Observatory for the kind hospitality. During
part of this work, IN was supported by an external ESA fellowship. This work is
partially supported through CNAA, ASI and MURST grants.
We also thank the anonymous referee for her/his quick reply and useful
comments, suggestions and corrections that increased the readability  of the
paper.
 \end{acknowledgements}


\end{document}